\documentclass{ws-mpla}
\usepackage[super]{cite}
\usepackage{graphicx,float}
\usepackage{hyperref}
\usepackage{cite}
\usepackage{xcolor}

\begin{document}

\markboth{Yasaman K. Yazdi}{Everything You Wanted to Know About CST and Cosmology}

 \title{Everything You Always Wanted to Know About How Causal Set Theory Can Help with Open Questions in Cosmology, but Were  Afraid to Ask}


\author{Yasaman K. Yazdi\footnote{present address}}

\address{School of Theoretical Physics, Dublin Institute for Advanced Studies,\\ 10 Burlington Road, Dublin 4, Ireland.\\
ykyazdi@stp.dias.ie}

\maketitle

\begin{abstract}
We briefly survey open questions in cosmology that either have been or could potentially be addressed with the help of the causal set theory approach to quantum gravity. Our discussion includes topics ranging from dark matter and dark energy to primordial quantum fluctuations and horizon entropy. By putting together all in one place these cosmologically relevant directions of work within causal set theory, we hope to impart a bird's-eye view of this important research theme.
\end{abstract}

\vspace{2pc}

\today
\vspace{0.75cm}
\section{Introduction}

Quantum gravity, as the name suggests, aims to understand the quantum nature of gravity. At its most basic level, this means that we would like to know how gravity behaves at the smallest of length scales, or largest of energies (close to the Planck scale), where our current theories of it break down. This breakdown occurs for example in spacetime singularities of classical general relativity, such as at the big bang singularity.

As important as this most basic aim is, to understand the physical nature of gravity all the way to the Planck scale, quantum gravity promises to do a lot more. There are many other important open questions in modern theoretical physics which quantum gravity is expected to give an answer to. Prominent examples include the black hole information paradox and the cosmological constant problem. 

There are also a multitude of additional open questions (e.g. cosmological tensions and quantum foundational puzzles),  which quantum gravity \emph{may} give answers to. There is no consensus regarding which (if any) of these auxiliary questions an approach to quantum gravity must shed light on. Different communities of researchers have different thoughts on this, and different approaches to quantum gravity have different capabilities to tackle these additional open questions. What there is consensus on, however, is that we welcome a fundamental quantum gravitational understanding of any number of these open questions.

The focus of this review is open questions in cosmology, and how ideas from the causal set theory approach to quantum gravity may help to answer them. As we will discuss below, causal set theory has already had some successes in addressing cosmological open questions. It also has something to offer to the auxiliary cosmological questions such as the recent Hubble tension.

We begin in Section \ref{sec:cst} with an introduction to causal set theory. In Section \ref{sec:early} we discuss questions from the early universe, including what happens at the beginning and the origin of primordial quantum fluctuations.  In Sections \ref{sec:dm} and \ref{sec:elam} we review candidates for dark matter and dark energy respectively. Finally, in Section \ref{sec:entropy} we summarize several approaches to understanding the microscopic origin of horizon entropy. We end with some concluding remarks in Section \ref{sec:conclusion}.

\section{Causal Set Theory}\label{sec:cst}
\begin{figure}[h]
    \centering
    \includegraphics[width=.3 \linewidth]{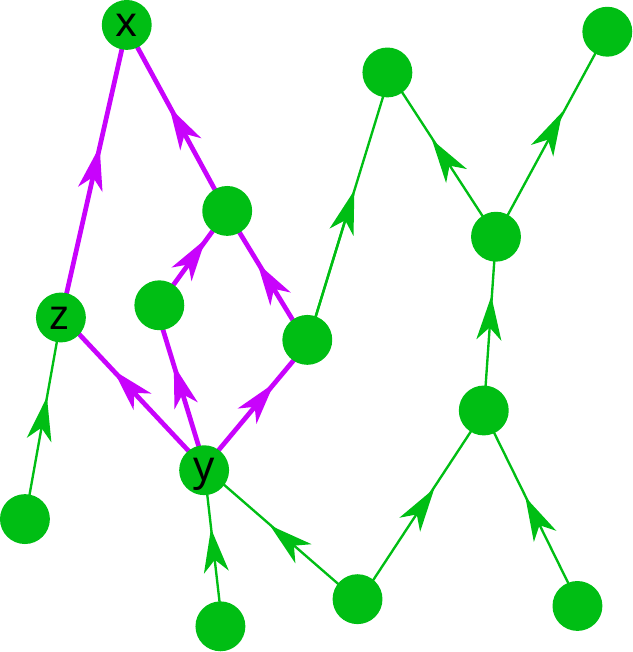}
    \caption{A Hasse diagram of a small causal set.  Lines indicate links (i.e., a causal relation between two elements with no element in between) and the arrow indicates the direction of the causal relation.  Elements not directly linked can still be related by transitivity.  For example in the diagram $y\prec z$, $z\prec x$, and therefore $y\prec x$.  Finally, the elements joined by purple lines form a {\it causal diamond} with endpoints $x$ and $y$. A causal diamond or causal interval between $x$ and $y$ is the intersection of the causal future of $y$ and the causal past of $x$.}
    \label{fig:causet}
\end{figure}
Causal set theory\cite{sorkin1991spacetime,bombelli1987space,surya2019causal} is an approach to quantum gravity in which a continuum  spacetime, described by a Lorentzian manifold and metric $(\mathcal M, g)$ in general relativity,  is replaced by a fundamentally discrete causal set and partial order $(\mathcal C, \prec)$. The partial order $\prec$ describes the causal relations among the discrete elements and has the same properties as the causal structure in general relativity.  For example, $z\prec y$ and $y\prec x$ indicate that $y$ and  $z$ (the latter by transitivity of the relation $\prec$) causally precede $x$. See Figure \ref{fig:causet} for an example of a small causal set. This discreteness is a natural solution to the infinities that arise in quantum field theory as a result of arbitrarily short distances or high energies. Discreteness also eliminates the spacetime singularities of general relativity. 

A continuum spacetime can resemble certain discrete causal sets if it  does not have any features below the discreteness scale. When this is the case, the correspondence between the continuum spacetime and discrete causal set is in terms of a number-volume correspondence: there are roughly equal numbers $N$ of causal set elements found embedded in equal generic volumes $V$ of the continuum spacetime, up to a proportionality constant $N\sim \rho V$. Throughout this review we will work in fundamental units where $\rho=1$. 
\begin{figure}[h]
    \centering
    \includegraphics[width=.5 \linewidth]{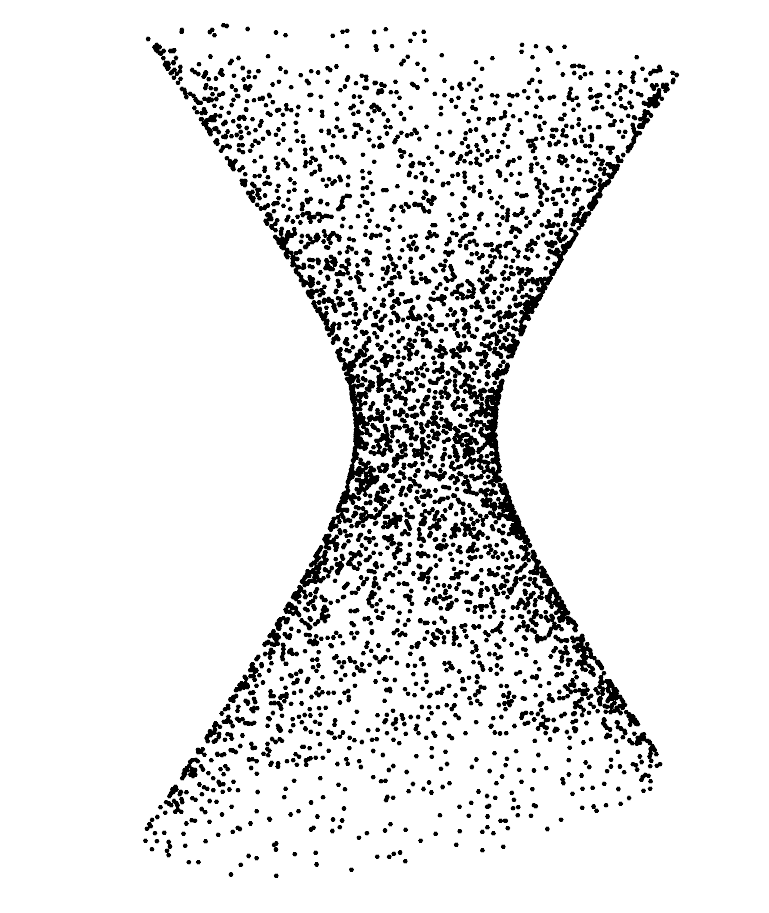}
    \caption{A 5000 element sprinkling into an interval in $1+1$D de Sitter spacetime. The elements are positioned according to their embedding in one higher dimension Minkowski spacetime.}
    \label{fig:desitter}
\end{figure}
While the dynamics\footnote{In the next section we will discuss an approach to the classical dynamics.  Most efforts in the quantum direction involve the path integral or sum over histories approach in one way or another.} (which is under development) must ultimately explain how such a correspondence arises, at the kinematic level we can ensure this correspondence via a process called Poisson sprinkling: we sample or sprinkle points randomly and uniformly in $\mathcal M$ such that the number of points in an arbitrary and generic (sub)volume follows a Poisson distribution. This randomness of the Poisson distribution is key for achieving a distribution where the number of elements within each generic volume or subvolume is statistically proportional to the volume. We then endow the elements with the same causal relations as their corresponding points in the original spacetime. The resulting causal sets have the important feature that they are locally Lorentz invariant. No frame is selected by the random Poisson distribution of elements, as desired for an underlying quantum gravitational structure. Figure \ref{fig:desitter} displays an example sprinkling into a finite volume interval of de Sitter spacetime.

\section{Early Universe}\label{sec:early}
\begin{figure}[h]
    \centering
    \includegraphics[width=.75 \linewidth]{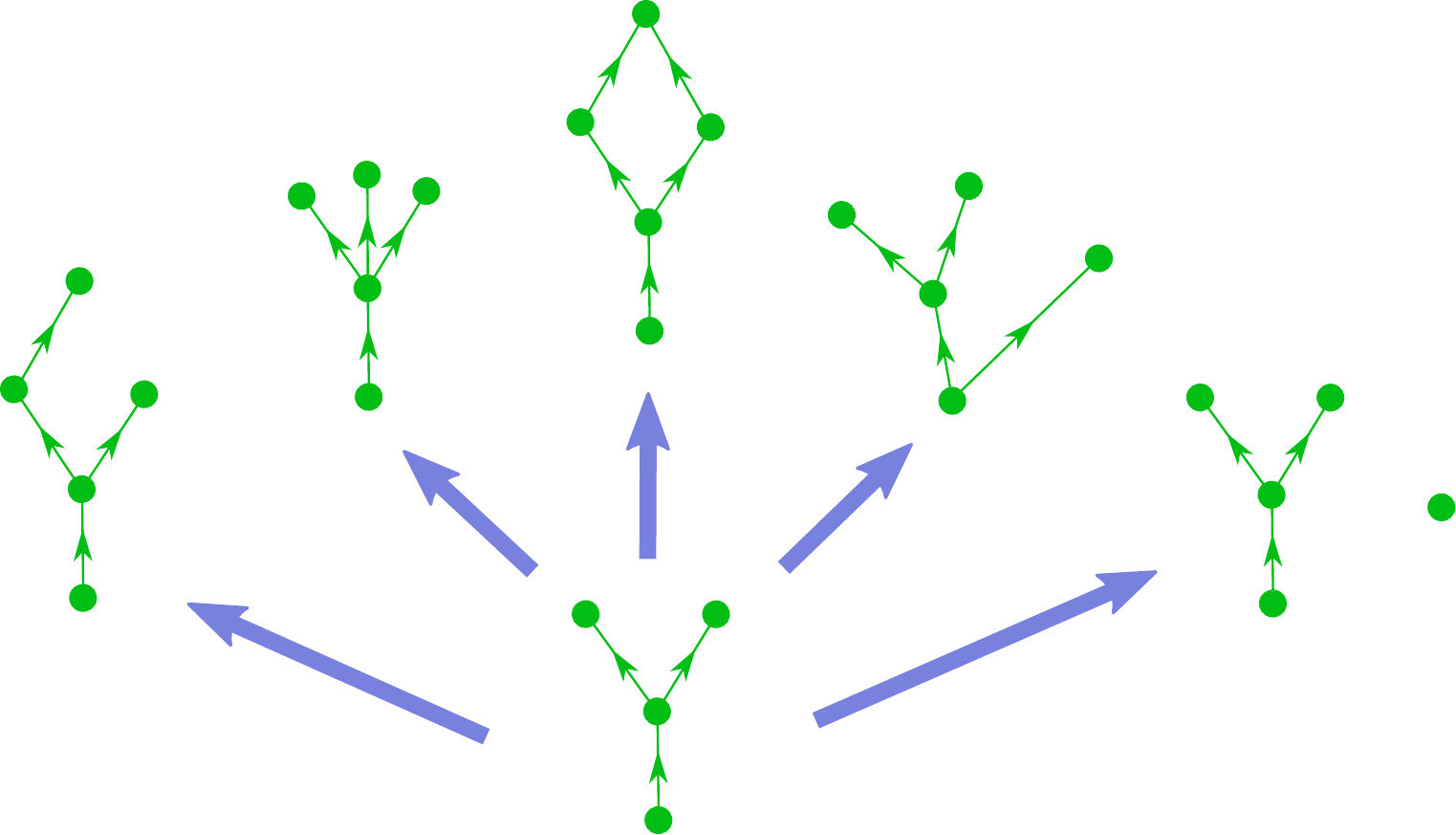}
    \caption{One step in a CSG process.  The causal set of size $4$ at the bottom can grow into one of the five causal sets above.}
    \label{fig:csg}
\end{figure}

There has been some progress in developing classical dynamical models for causal sets which have features resembling the early universe. A family of these models are known as ``classical sequential growth" or CSG models.\cite{Rideout:2001gu} As the name suggests, in these models, a causal set  grows element by element sequentially. The sequence of births of elements is merely a gauge that facilitates describing the process. At stage $1$, the first element is born, and subsequently at each stage $n$ we have an $n$-element causal set. Each new $n^{\text{th}}$ element selects a random subset of the existing elements to be in its causal past. While this choice is made randomly or stochastically, there are some rules to ensure that causality and a suitable analog of discrete covariance are preserved by the ancestor selection process. Figure \ref{fig:csg} shows one step in a CSG process.

Below we discuss some of the interesting cosmological implications for the early universe that result from classes of these CSG models. 
\subsection{Initial Singularity}
How did our universe as we know it begin? The big bang is the conventional answer to this question provided by general relativity. Of course the big bang singularity represents a regime where the concept of spacetime ceases to be meaningful. Causal sets, by virtue of their fundamental discreteness, evade such singularities. The minimum separation between the elements (presumed to be close to the Planck length) tames infinite curvatures and energies.

The early causal sets at the very early universe will be  non-geometric\footnote{In the limit of the number of elements $N\gg 1$ the notion of a continuum spacetime, for example one that can be approximated via a sprinkling, becomes meaningful.}, but they will nevertheless have features we have come to expect from geometric scenarios. For example, a broad class of CSG models lead to an early structure that mimics the big bang. This big bang analog is a causal set element that causally precedes every other element in the causal set. This original element is an example of what is known as a \emph{post} (an element causally related to all the other elements, be it a past or future relation). 
As we will discuss in the next subsection, a large class of models admitting such posts are followed by periods of rapid growth, which further resembles the behavior close to the big bang.

\subsection{Early Exponential Expansion}
The first moments after the big bang are believed to consist of a period of exponential expansion. This is the best explanation we have for observations such as the large size of the universe, high degree of isotropy, homogeneity, spatial flatness, and the lack of monopoles. However, the origin of this explanation requires explanation itself, and this is a current open question. Inflation\cite{guth,Liddle:1999mq,Tsujikawa:2003jp} is a popular and practical model (based on the existence of a scalar field(s) with various possible potentials) that is designed to admit such a period of rapid expansion. However, inflation is emphatically not an example of the kind of fundamental understanding, mentioned in the introduction, that we wish to obtain of our universe. For example, not only does it not stem from first principles, the framework of cosmic inflation is too flexible, making it difficult to pinpoint what could be direct evidence for it.\cite{pop, earman_mosterin_1999} Inflation may in the future find its missing motivations if a theory of quantum gravity predicts a particular kind of it.

A similar rapid early growth as our universe occurs in one of the simplest families of sequential growth dynamics called \emph{transitive percolation}\cite{Rideout:1999ub,Rideout:2000fh, Rideout:2001gu}. These models have one free parameter $p$; at each stage $n$ of the growth process, the new element picks with probability $p$ each existing element to be in its past. An $n$-element causal set grown by such a dynamics can be modelled using the following algorithm: If we have labelled the elements $1, 2, ..., n$, the probability to introduce a relation $i\prec j$ between any pair of elements $i\in \{1,2,...,n-1\}$ and $j\in \{i+1,i+2,...,n\}$ is $p$. We must also ensure transitivity and add all the relations that are implied by the new relations through transitivity. 

\emph{Originary percolation} is a version of transitive percolation with the condition that each new element needs to be related to at least one other existing element.\cite{Varadarajan:2005gg} These models ensure that the universe begins with a post resembling the big bang, as discussed in the previous subsection. 

It is known that for small $p$, originary percolation leads to a tree-like growth until $n\sim 1/p$ elements have accumulated. This tree-like structure, which resembles a rapid growth, prompted  investigations into how closely there might be a similarity to a spacetime such as  de Sitter. \cite{Ahmed:2009qm} Despite the non-geometric nature of these early causal sets, proper time and spacetime volume estimates are still well-defined  and can be used for comparisons with continuum geometries. The causal set analog of proper time between two (causally related) elements is the length of the longest chain between them.\cite{brightwellt} A chain between arbitrary elements $x_1$ and $x_n$ is a totally ordered sequence of (any number of) relations between them that ultimately relate them by transitivity: $x_1\prec x_2\prec...\prec x_n$. The spacetime volume of the causal diamond between two elements can also be approximated by counting the number of elements within it. In Ref. \refcite{Ahmed:2009qm} these analogs of proper time and spacetime volume were used to investigate whether there was a statistical resemblance to de Sitter spacetime (of any dimension). These studies did find a resemblance to de Sitter spacetime\footnote{Some parts of their analysis favored $2+1$ dimensions, while the majority favored $3+1$ dimensions.} after the tree-like era of the percolation dynamics.

Another known feature of transitive percolation is that it describes a cyclic cosmology, i.e. one that undergoes alternating periods of expansion and contraction.\cite{Sorkin:1998hi}  More general sequential growth dynamics that lead to cyclic cosmologies have also been studied in some detail.\cite{Dowker:2022ehl, Sorkin:1998hi, Martin:2000js}  In some of these models, the dynamics of each new cycle is different from the previous one which we can interpret as a kind of cosmological renormalization group.\cite{Martin:2000js} A major open question is whether and how one of these dynamics eventually evolves into a manifoldlike causal set such as one approximated by a Friedman-Lema\^itre-Robertson-Walker (FLRW) universe.\cite{Rideout:2000fh, brightwell} A hope would be that the emergence of manifoldlike behaviour will be better understood once the quantum dynamics is developed. Some indications of how this may happen are given in Ref. \refcite{Carlip:2022nsv}, where a suppression of a large class of non-manifoldlike causal sets is demonstrated in a path sum.

A last remark on this topic is that such rapid early expanding behavior has also been observed in two other independent contexts. One is Ref. \refcite{Glaser:2014dwa}
 where the Hartle-Hawking wavefunction in two dimensions was studied using the Benincasa-Dowker causal set action\cite{Benincasa:2010ac}. The other context is when a positive initial $\Lambda$ is set for  Everpresent $\Lambda$ dark energy models, discussed in Section \ref{sec:elam}.\cite{Das:2023hbw}

\subsubsection{Primordial Quantum Fluctuations}\label{sj}
One of the successes of quantum field theory in curved spacetime has been to generate a primordial power spectrum that closely agrees with the observed anisotropies in the cosmic microwave background. The detailed nature of these fluctuations and how they lead to structure formation is still not well understood. Many models for these primordial fluctuations use some form of (quantum) inflaton field.\cite{mukhanov} As mentioned above, we would like to have a model that naturally connects to a more fundamental theory, and so far that has been challenging to achieve with inflation. In this subsection, we comment on some attractive features of studying quantum field theory on background causal sets approximated by curved spacetimes. These comments apply for general scalar fields, irrespective of whether or not they are an inflaton field.

In quantum field theory in curved spacetime, the background spacetime is kept classical while the field theory living on it is quantum, and the spacetime and field are assumed not to backreact onto one another. An understanding of the effects of backreaction must await a full understanding of quantum gravity. Quantum field theory on a background causal set can be viewed as an increment closer to quantum gravity in comparison to regular quantum field theory on curved spacetime. 

Operator expectation values are important physical quantities in quantum field theory. In many cases, the entire physical content of the theory is captured by the set of n-point correlation functions of the field operator.\cite{Chen:2020ild} These expectation values require a notion of \emph{state} to compute. Choosing a state in arbitrary curved spacetimes is a notorious challenge as there is no unique or obvious choice, and different choices lead to different consequences. Since cosmological spacetimes are examples of curved spacetimes, they face the same challenge of an appropriate choice of state. The Sorkin-Johnston (SJ) state\cite{Johnston:2009fr, Sorkin:2017fcp} is a choice of vacuum state in any arbitrary curved spacetime that is a candidate for circumventing this challenge. The SJ state is unique and distinguished in any globally hyperbolic spacetime and typically reflects the symmetries of the background spacetime (if there are any). The SJ state is also well-defined for both causal sets and continuum spacetimes, but with the added advantage in causal sets that the field is automatically UV-regulated. Let us briefly review the definition of this state and the quantum field theory that ensues:

We will focus on the free field theory, though extensions to the interacting case\cite{Chen:2020ild} have also been studied\refcite{}. Central to the  SJ field theory is the Pauli-Jordan function or spacetime commutator 
\begin{eqnarray}
i \Delta  (x,x '  ) &=&\langle 0| [ \phi  (x  ),\phi  (x'    )   ]|0  \rangle\nonumber\\
 &=& {W } (x,x '  )-{W }(x',x  ),
\end{eqnarray}
where $x$ is a spacetime point and $W$ is the Wightman function or spacetime two-point correlation function. 
$i \Delta $ is a c-number and can also be expressed using the  retarded and advanced Green functions ($G_{R}$ and $G_{A}$ respectively) as: $\Delta(x,x')=G_{R}(x,x')-G_{A}(x,x')$. With $i\Delta$ or $G_R$  as the starting point, the SJ  prescription is manifestly  covariant. $i \Delta $ defines an integral operator over the Hilbert space of square integrable functions ${L}^{2}(R)$ in a region of spacetime $R$. Under  suitable conditions (which are satisfied in finite volumes of spacetimes such as de Sitter and FLRW) in  globally hyperbolic spacetimes or subregions, $i\Delta$ is self-adjoint and can be decomposed into its positive and negative eigenspace.
 
Let $\{u_{k}\}$ and $\{v_{k}\}$ be the set of normalized positive and negative eigenfunctions of $i\Delta$ respectively and $\pm\lambda_k$ their corresponding eigenvalues. The SJ Wightman function (and therefore state definition) is obtained by taking the  positive eigenspace of $i\Delta$:
\begin{equation}
{W}_{SJ}(x,x')\equiv  \text{Pos}(i\Delta)=\sum_{  k}^{ }\lambda _{k}u_{k}(x)u_{k}^{\dagger}(x').
\end{equation}
In a causal set $x,x'$ are elements of the causal set, and since there are a finite number $N$ of them in a finite region, $i\Delta$ and $W_{SJ}$ are finite $N\times N$ dimensional matrices. This finiteness also turns the  integral equations resulting from the continuum integral operator $i\Delta$ into simpler algebraic operations in the causal set case. The SJ state has been studied in causal sets sprinklings of de Sitter spacetime\cite{Aslanbeigi:2013fga, Surya:2018byh} and continuum FLRW \cite{Afshordi:2012jf}. In the de Sitter case, a range of masses were studied in $3+1$D and the vacua were shown to be de Sitter invariant, though not always corresponding to an $\alpha$-vacuum.\cite{Surya:2018byh} Therefore, in some cases, new vacua were obtained. In the work in FLRW, correlations on super-horizon scales were found, without the help of a phase of rapid (inflation-like) expansion.\cite{Afshordi:2012jf} It would be interesting to explore the phenomenological consequences of these findings regarding the SJ state.

There is also a recent direction of work that has made predictions for a signature of a covariant UV cutoff at the level of quantum field theory in curved spacetime.\cite{Chatwin-Davies:2022irs} Many of the ingredients of this work (e.g. Green functions and d'Alembertians) are available in causal set theory as well, in addition to it being a concrete example of a theory with an intrinsic and covariant UV cutoff. It would be interesting to explore whether these predictions can be realized or sharpened in the context of causal sets.

\section{Dark Matter}\label{sec:dm}
The origin of dark matter, the mysterious dominant mass density in our universe, is a major open question in cosmology. Studies of nonlocal quantum fields on causal sets led to a surprising candidate for dark matter which became known as off-shell dark matter (O$f$DM) ~\cite{Saravani:2015rva, Saravani:2016enc, Saravani:2019smt}. See also Ref. \refcite{pirsa_Niayesh} for a summary of this work.

Nonlocal analogues of the d'Alembertian operator $\Box$ that describe the propagation of scalar fields, exist more commonly in causal set theory than their more conventional local counterparts. Let us recall what one such  discrete nonlocal operator $B$ and its continuum limit $\Tilde{\Box}$ look like. Our discussion will be limited to scalar field theories. The extension to other kinds of fields is not presently known.

An explicit example of one of these nonlocal evolution operators in $3+1$D is \cite{Benincasa:2010ac}
\begin{equation}
B\phi(x)=\frac{4}{\sqrt{6}}\left(-\phi(x)+\left(\sum_{i=0}^3c_i\sum_{y\in \lozenge_i}\right)\phi(y)\right),
\label{box2}
\end{equation}
where $y\in \lozenge_i$ means that $y\prec x$ and the cardinality of the causal diamond between $x$ and $y$ (not including $x$ and $y$ themselves) must be $i$. The coefficients $c_i$ in this case are: $c_0=1, c_1=-9, c_2=16, c_3=-8$. The continuum average\footnote{By continuum average rather than continuum limit we mean that at a fixed causal set density and for fixed $x$ we consider an ensemble of sprinklings and compute $B\phi(x)$ for each sprinkling in this set. The average of the $B\phi(x)$ over all these sprinklings is $\Tilde{\Box}\phi(x)$.} of \eqref{box2} is
\begin{equation}
\Tilde{\Box}\phi(x)=\frac{4}{\sqrt{6}}\left(-\phi(x)+\sum_{i=0}^3\frac{c_i}{i!}\int_{J^-(x)}e^{-V_{xy}}\,(V_{xy})^i\phi(y)\,d^4y\right),
\label{box2_cont}
\end{equation}
where $V_{xy}$ is the volume of the causal diamond between $x$  and $y$. At low energies and in the continuum limit, these nonlocal d'Alembertians reduce to the usual local one in Minkowski spacetime. In curved spacetimes, the continuum limit of  these nonlocal d'Alembertians contains a term proportional to the Ricci scalar curvature in addition to the local d'Alembertian.\cite{Benincasa:2010ac} Note that \eqref{box2} and \eqref{box2_cont} are nonlocal because in order to know $B\phi(x)$ at $x$, we must have information beyond $x$, i.e., we must know the  elements $y$ to the past of $x$. However, this  nonlocality has a spacetime (more specifically, retarded and causal) rather than spatial nature. \eqref{box2} is just one example of a nonlocal d'Alembertian operator arising in causal set theory. Entire classes of  these operators have been defined and studied. We refer the reader to Refs. \refcite{Benincasa:2010ac} \refcite{Belenchia:2014fda}\refcite{Belenchia:2015hca} \refcite{Aslanbeigi:2014zva} \refcite{Dowker:2013vba} \refcite{Sorkin:2007qi} for more information regarding this work including some of its  non-cosmological phenomenological implications.

The authors of Ref. \refcite{Saravani:2015rva} used the Schwinger-Keldysh double path integral approach to study the quantization of a free massless scalar field with classical evolution described by nonlocal d'Alembertians such as \eqref{box2_cont}. They found a set of on-shell and off-shell solutions in the mode expansion of the quantized field operator. These off-shell particles do not satisfy the conventional dispersion relation. Interestingly, when studying scattering processes in the interacting theory, they found that the on-shell quanta can scatter and produce off-shell particles, but that the off-shell particles do not (non-gravitationally) interact. In other words,  any scattering with off-shell particles has zero cross-section. This prompted them to consider these off-shell particles as a natural candidate for dark matter. 

Since the production rate of the off-shell particles is suppressed by the nonlocality scale (presumed to be close to the Planck scale), the phenomenological picture is that the majority of dark matter was produced in the early universe. Specifically, post-inflation reheating and radiation self-interaction provide the dense and hot conditions that lead to efficient relevant scattering processes.

An attractive feature of this proposal is that it suggests that dark matter could be related to known matter. This is in contrast to other models that introduce new and exotic particles. It also suggests a reason why dark matter exists and why it is so difficult to detect. Despite this detection challenge, in Refs. \refcite{Saravani:2016enc} and \refcite{Saravani:2019smt} some potentially observable signatures of this model are suggested.

\section{Dark Energy}\label{sec:elam}
Like dark matter, the nature and origin of dark energy is a big mystery in cosmology. Everpresent $\Lambda$ is a dynamical dark energy model that naturally follows from basic ingredients in causal set theory. A recent comprehensive review, status report, and list of references on this topic can be found in Ref. \refcite{Das:2023hbw}. Below we will briefly review the assumptions that underlie the idea, before discussing the open questions it is relevant for. 
\subsection{Everpresent $\Lambda$}
There are three main assumptions that form the basis of the Everpresent $\Lambda$ heuristic argument:

\paragraph{Assumption 1}: Causal sets that macroscopically resemble continuous spacetimes (e.g. the FLRW universe) are those produced by a Poisson sprinkling. This in turn tells us that if we have a region of spacetime with fixed volume $V$, according to the Poisson distribution, we should expect to have a mean number of $\langle N\rangle=V$ elements in that volume, with standard deviation $\Delta N=\sqrt{N}$. Turning this correspondence around, if we have a causal set with a fixed number $N$ elements, there will be an ensemble of continuum spacetimes (or spacetime regions) that it can resemble, with mean spacetime volume $\langle V\rangle=N$ and standard deviation
\begin{equation}
\Delta V=\sqrt{V}.
\label{volstd}
\end{equation}
\paragraph{Assumption 2}: Causal set dynamics involves the path integral and it is natural to hold the volume fixed as we perform the integral.\footnote{Traditionally, one is used to seeing the path integral applied over a fixed time interval. In a causal set, the passage of time is more naturally described by the number of elements which accumulate as the causal set grows through a dynamical process. Translating back to the continuum via the $N\sim V$ correspondence, holding the volume fixed would be the natural constraint in an effective gravitational path integral coming from causal set dynamics.} This is analogous to the constraint considered in unimodular theories of gravity. This volume fixing in the path integral leads to a  conjugacy between the cosmological constant $\Lambda$ and the spacetime volume $V$.\footnote{The close relation between these two quantities can already be seen from how they appear multiplying each other in the gravitational action with a cosmological constant term.} The corresponding uncertainty relation between them is
\begin{equation}
\Delta V\Delta\Lambda\sim 1
\label{uncertainty}
\end{equation}
or with a slight rearrangement, $\Delta\Lambda\sim 1/\Delta V$, in units where $\hbar=1$. \\

\paragraph{Assumption 3}: The mean value of $\Lambda$ about which the fluctuations occur is zero. Hence we have 
\begin{equation}
\Lambda\sim\Delta\Lambda.
\label{mean}
\end{equation}
More precisely, in writing \eqref{mean} we really mean that we expect $\Lambda\in[-\Delta \Lambda,\Delta \Lambda ]$. In other words it is natural for the cosmological constant $\Lambda$ to take any value in the interval $[-\Delta \Lambda,\Delta \Lambda ]$ since this is roughly the standard deviation interval. 

This third assumption is the only input to the Everpresent $\Lambda$ proposal that does not stem from a fundamental aspect of causal set theory. 

\begin{figure}
    \centering
    \includegraphics[width=1 \linewidth]{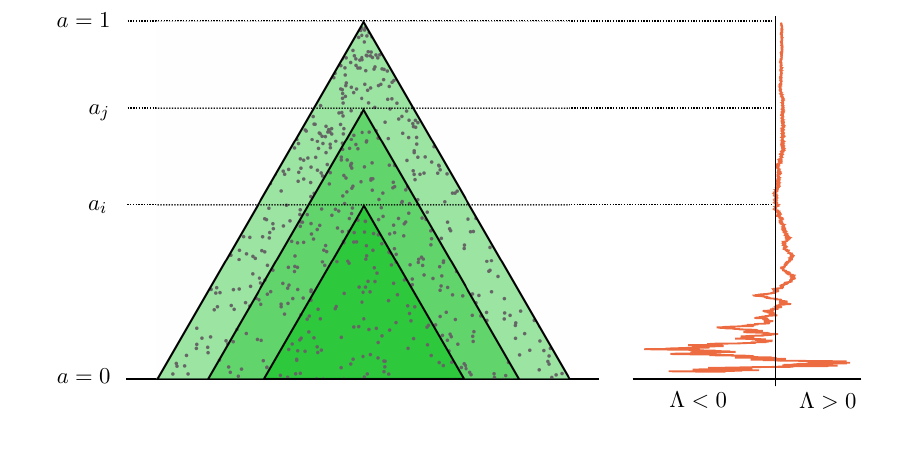}
    \caption{On the left is the past lightcone of an observer at different times (indicated by the conformal factor $a$).  On the right is the value of $\Lambda$ obtained from an Everpresent $\Lambda$ simulation.  Note how the fluctuations of $\Lambda$ decrease as the volume of the past lightcone increases, in agreement with \eqref{lamV}.}
    \label{fig:everlam}
\end{figure}

Putting together \eqref{uncertainty} and \eqref{mean} we have $\Lambda=\Delta\Lambda\sim1/\Delta V$. Finally substituting the variation of $V$ from \eqref{volstd} we get
\begin{equation}
    \Lambda\sim 1/\sqrt{V}.
\label{lamV}
\end{equation}
The procedure above still leaves open the question of what volume to insert in \eqref{lamV}. A \emph{causal} choice would be to take the volume of our past lightcone, and this is indeed what is considered in the literature on Everpresent $\Lambda$.  See Figure \ref{fig:everlam} for an example of a model based on this and the corresponding time dependence of $\Lambda$. The spacetime volume of the observable universe is approximately $10^{244}$ Planck volumes. Using this as our estimate for the volume of our past lightcone today leads us to $\Lambda\sim 10^{-122}$. Thus we have remarkably arrived at the rough value of the observed cosmological constant today without any fine tuning. This is the resolution of the cosmological constant problem that  Everpresent $\Lambda$ offers. The heuristic of Everpresent $\Lambda$ was initially introduced in Ref. \refcite{Elam_og}. Other early references include Refs. \refcite{DOlivo:1991hpg}. Notice that in estimating the spacetime volume of the observable universe as $\sim10^{244}$ Planck volumes, we have made the assumption that we live in four spacetime dimensions. If we had assumed a higher or lower spacetime dimension we would end up with a smaller or larger (respectively) volume and hence a cosmological constant that is typically too large or too small (respectively) compared to the observed value. Hence, if the Everpresent $\Lambda$ picture is correct, it lends evidence and insight to the fact that we live in four spacetime dimensions.

There are additional cosmological tensions that Everpresent $\Lambda$ can potentially help with. Below, we discuss two well known ones.

\subsection{The Hubble Tension}
The Hubble tension is the name given to a clash in 
recent measurements of the Hubble parameter $H_0$, with disagreements between supernova (together with Cepheid stars and tip of the red giant branch) data and observations from the cosmic microwave background. Both sets of measurements are robust, leaving little room for systematic error to account for the discrepancy. This disagreement forms the basis of the Hubble tension \cite{Kenworthy:2019qwq, Efstathiou:2020wxn, Abdalla:2022yfr}. In light of this disagreement, many believe that we must abandon traditional models where  $\Lambda$ is a constant and instead consider dynamical models where it can have an evolving value, thereby relieving the tension. Everpresent $\Lambda$ is precisely such a dynamical model. In particular, early dark energy models where there is extra dark energy in the period after matter-radiation equality but before recombination, are a leading candidate to alleviate this tension.\cite{Poulin:2018cxd,Abdalla:2022yfr,  Das:2023rvg} Due to its random fluctuating nature, such a period of early dark energy can be realized with Everpresent $\Lambda$.

\subsection{The $S_8$ (or $\sigma_8$) Tension}
Recent observations of large-scale structure have set stringent constraints on the clustering of matter. This has revealed a tension in the strength of clustering inferred from different probes.\cite{Abdalla:2022yfr} $S_8\equiv \sigma_8\sqrt{\Omega_m/0.3}$ is the parameter used to quantify this tension, where $\Omega_m$ is the fraction of the total energy density belonging to matter. In particular, CMB data favors a higher  value of $S_8$ while  lower redshift data such as from weak lensing and galaxy clustering prefer a lower value of $S_8$.

Models in which the cosmological constant changes sign over the history of the universe have been shown to relax the $S_8$ tension.\cite{Akarsu:2021fol, Akarsu:2022typ} Since Everpresent $\Lambda$ is likely to change sign over the history of the cosmos, it can behave like these models and potentially relax the $S_8$ tension in the same manner.

Another class of models that have been shown to ameliorate the $S_8$ tension are interacting dark energy models where the energy density can be transferred between dark matter and dark energy.\cite{Lucca:2021dxo, Poulin:2022sgp} While this scenario does not automatically occur in current Everpresent $\Lambda$ models, it may be an interesting avenue to modify the present models to incorporate such an interaction.

At present, there are only qualitative indications that Everpresent $\Lambda$ possesses some of the features that are believed to be key to resolving the Hubble and $S_8$ tensions.  Carrying out a dedicated and quantitative study of how much the current Everpresent $\Lambda$ models (or slightly modified versions of them) can alleviate these tensions is one of the most important directions of future work on Everpresent $\Lambda$. The $S_8$ and Hubble tensions are correlated tensions, namely, many proposed models that alleviate one tension tend to worsen the other. Therefore it would be very interesting to see if Everpresent $\Lambda$ can alleviate both at the same time. 

\section{Cosmological Horizon Entropy}\label{sec:entropy}
In 1974 Hawking showed that when quantum effects are taken into account, black holes are not so black after all. In fact they radiate like a blackbody at a characteristic temperature related to their surface gravity\cite{HawkingRadiation}. Around the same time, a characteristic entropy proportional to the area of the event horizon of a black hole was also discovered. This entropy is now called the Bekenstein-Hawking black hole entropy \cite{Bekenstein}. 

Not long after, these thermodynamic properties were extended to cosmological horizons (in models with a positive cosmological constant) as well.\cite{gibbons} Similar to the black hole case, these cosmological event horizons were also found to have a temperature and an  entropy proportional to their area associated with them.

The derivations of these thermodynamic phenomena require arguments from both classical gravity and quantum theory, making them intrinsically “quantum gravitational”. They have thus been key  reference points in the search for a theory of quantum gravity.  Motivated by how in ordinary thermodynamics, thermal properties are described by the statistical mechanics of the underlying microstates, an active field of research has been to try to understand these effects from a statistical mechanical perspective \cite{Bombelli, Rovelli, Witten, Strominger:1996sh, Mathur:2005zp, Maldacena:1998bw}. 
Research in causal set theory has also been active in this direction. In this section we will focus on the outstanding question of: what is the microscopic origin of these (cosmological) horizon entropies?\cite{Sorkin:1997ja}. There have been two distinct approaches to try to answer this and we discuss each of them below.

\subsection{Entanglement Entropy}
The entanglement entropy of a quantum field confined to  a spacetime subregion, often scales as the spatial area of the subregion. This ``area law'' is what makes entanglement entropy a natural candidate to be the microscopic source of horizon entropy. It is also then no surprise that the concept of entanglement entropy sprung from the search for a deeper understanding of black hole entropy \cite{Sorkin:1984kjy, Bombelli}. 

\begin{figure}
    \centering
    \includegraphics[width=.75 \linewidth]{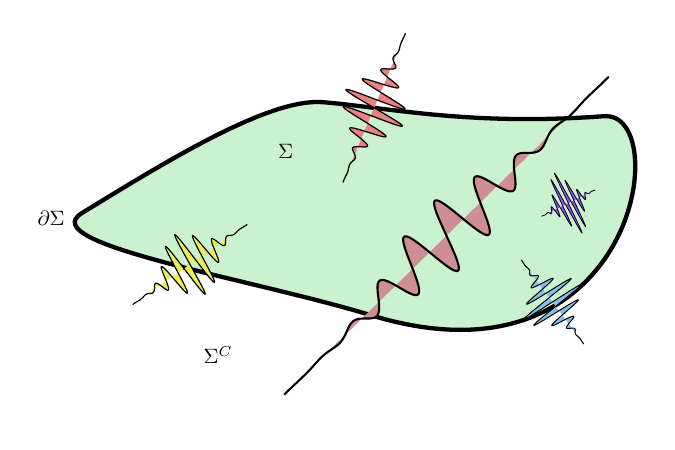}
    \caption{A sketch of the contributions to entanglement entropy.  The field modes that straddle the boundary $\partial\Sigma$ of the (spatial) region $\Sigma$, contribute to the entanglement entropy of the field in the region.  Modes that do not straddle the boundary, e.g. the mode completely contained in $\Sigma$, do not contribute to the entanglement entropy.  
    A UV cutoff (e.g. a minimum wavelength) is needed, otherwise the modes straddling the boundary can become arbitrarily short, and  an infinite number of them would contribute to the entanglement entropy.\\ \\
    }
    \label{fig:ee}
\end{figure}

The entanglement entropy of a quantum field described by (an initially pure) density matrix $\rho$ is 
\begin{equation}\label{ee}
      S = -\text{Tr} \rho_{\text{red}}\ln\rho_{\text{red}},
\end{equation}
where $\rho_{\text{red}}$ is the reduced density matrix encoding the field degrees of freedom in the confined subregion (e.g. the region of spacetime on one side of a horizon) only, or alternatively in its causal complement only.\footnote{By the complementarity property of entanglement entropy, we would get the same answer in either case.
} Intuitively, \eqref{ee} quantifies how much information one loses as a result of the limited access to the quantum field. See Fig. \ref{fig:ee} for a sketch of a spatial version of an entangling region. In some sense, the entanglement entropy counts the number of field modes between the inside and outside of an entangling region, straddling the boundary.  Some form of UV cutoff must be taken into account when describing the degrees of freedom; in the absence of such a cutoff, one would end up concluding that an infinite amount of information is lost as a result of the confinement (corresponding to information encoding arbitrarily high energy degrees of freedom of the quantum field theory near the boundary). There are a variety of ways to compute \eqref{ee} and a variety of choices of UV cutoffs to obtain a finite and thereby meaningful entanglement entropy. In causal set theory, the entanglement entropy is expressed as a sum over solutions to a generalized eigenvalue problem involving the two-point correlation function of the theory, $W(x,x')$. The discreteness scale furnishes the covariant UV cutoff that makes this entropy finite. 

The eigenvalue problem that needs to be solved is
\begin{equation}
 W \, v = i\lambda \; \Delta \, v,  \            
\label{geneig}
\end{equation}
where $W$ (typically chosen to be $W_{SJ}$) and $i \Delta$ are finite-dimensional matrices that we saw earlier in Section \ref{sj}. The arguments $x$ and $x'$ of $W$ and $i \Delta$ must be restricted to lie in the subregion whose entanglement entropy we are interested in, before solving \eqref{geneig}. We must also ensure that the solutions to \eqref{geneig} satisfy $\Delta v\neq 0 $. The summation that yields the entropy in terms of the eigenvalues in \eqref{geneig} then is\footnote{This is the first formulation of entanglement entropy that admits the use of a covariant cutoff. This admission is a result of the use of \emph{spacetime} correlation functions in \eqref{geneig}.}
\begin{equation}
S=\sum_{\lambda} \lambda\ln |\lambda|.
\label{s4}
\end{equation}
 Two noteworthy advantages we gain by studying entanglement entropy in causal set theory are:

\paragraph{Universality.} Many attempts at deriving and understanding black hole thermodynamics are limited by special assumptions about the dimension \cite{Strominger:1996sh} and/or the asymptotics (e.g. the need for the spacetime to be asymptotically Anti-de Sitter \cite{Maldacena:1998bw}). Hence the techniques utilized and the insights gained lack universality. 
With causal sets we can study much more general spacetimes. In particular, any dimension and any asymptotics can be investigated, both in principle and in practice. Hence working in the context of causal sets gives us a broad and unified understanding of horizon entropy.

\paragraph{Background Independence.} Classical general relativity is covariant and there are tight constraints on Lorentz invariance violations \cite{Collins:2004bp}; therefore it is natural for quantum gravity to also be covariant. However, it is notoriously difficult to cast quantum theory in a frame-independent or background-independent language. Since defining the horizon entropy requires input from quantum theory, our current treatments of it inevitably involve a choice of background that influences our description of the state and degrees of freedom for the microstates. The formulation of entanglement entropy in terms of the spacetime correlators \eqref{s4} has paved the way for a fully covariant treatment of horizon entropy. Together with the SJ vacuum state, the application of this (entanglement) entropy definition in causal set theory in particular allows for the use of a fundamentally covariant UV cutoff.

For a recent summary of work on entanglement entropy in causal set theory, see Ref. \refcite{Yazdi:2022hhv}. Entanglement entropy as horizon entropy in a cosmological context was investigated in Ref. \refcite{Surya:2020gjj}. Specifically, causal sets approximated by $1+1$ and $3+1$ dimensional de Sitter spacetime were considered. To be more precise, finite slabs within de Sitter spacetime (rather than the infinite volume global dS) were considered, due to the finiteness of the causal set. The entanglement entropy resulting from the restriction to the event horizon within these slabs was then studied. There are some subtleties involving small eigenvalues of $i\Delta$ that need to be handled with care, after which the expected area laws are obtained. It would be interesting to revisit this problem in light of recent insights\cite{Keseman:2021dkf} into the aforementioned subtleties. Extensions to other dimensions and field theories can also be explored.

Whether or not the entanglement entropy of all fields in nature turns out to be the full story of horizon entropy remains to be seen. It is, however, at the very least an important part of the story.
\subsection{Horizon Molecules}
Another approach to understanding horizon entropy is motivated by considering the entanglement between the causal set elements  themselves, in the absence of any quantum matter fields. In analogy with a gas and the role played by its molecules to describe its entropy, the information content of a horizon is suggested to be encoded in suitably defined primitive constituents or ``horizon molecules''. These horizon molecules consist of linked elements with at least one element inside the horizon and at least one outside it. Assuming each molecule to contribute an equal amount to the horizon entropy, as a result of the link information across the horizon being lost, the horizon entropy is captured by counting the number of these  molecules.

Much like in the case of entanglement entropy we saw in the previous subsection, in the absence of a UV cutoff it is difficult to arrive at a finite entropy, as there would generically be an infinite number of horizon molecules in the continuum. It turns out that even in the causal set case one does not always obtain a finite number of any kind of horizon molecule, but there will be a finite number of particular kinds of horizon molecules.

\begin{figure}
    \centering
    \includegraphics[width=.7 \linewidth]{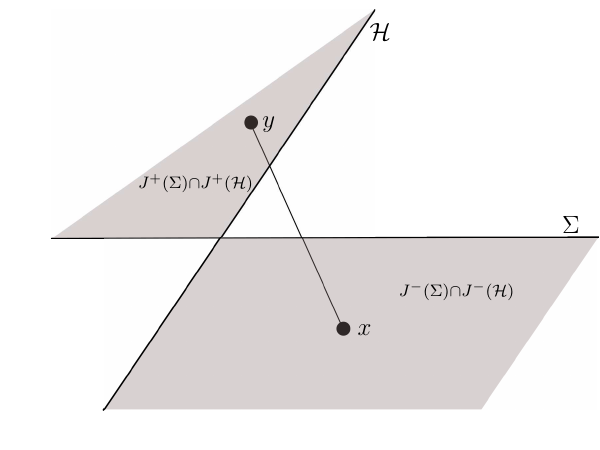}
    \caption{A link across an event horizon.  See equation \eqref{horizonmolecules}.}
    \label{fig:horizonmolecules}
\end{figure}

The original work on this topic considered $1+1$ dimensional black holes, including examples both in and out of equilibrium.\cite{Dou:2003af} Aiming for simplicity, a kind of link $x\prec y$ was chosen as the horizon molecule. These links consist of element $x$ outside the event horizon $\mathcal H$ and element $y$ inside, along with additional conditions which can be summarized as:
\begin{equation} \label{horizonmolecules}
\begin{split}
    x\in J^-(\Sigma)\cap {J}^-(\mathcal{H}) &~~\text{and is maximal in it}\\
    y\in J^+(\Sigma)\cap {J}^+(\mathcal{H})&\\
    x\prec y\wedge\nexists z| x\prec z\prec &y ~~\text{(i.e.}~ x\prec y ~\text{is a link)}\\
    y ~\text{is  minimal in}~ J^+(&\mathcal{H})
    \end{split}
\end{equation}
where $\Sigma$ is a null or spacelike hypersurface on which we are interested in evaluating the entropy. This hypersurface can be arbitrarily chosen and is a necessary reference in order to avoid overcounting the molecules. An element is maximal (minimal) in a causal set $\mathcal{C}$ if there are no elements to its future (past). In Ref. \refcite{Dou:2003af} counting of the molecules defined by the conditions above indeed led to a term proportional to the event horizon. Moreover, the coefficients were the same in both equilibrium and non-equilibrium cases, hinting at a universality.\footnote{This horizon molecules program also has the universality and background independence advantages highlighted in the previous subsection.} While in the absence of the quantum dynamics for causal sets, this treatment remains classical, it serves as an interesting derivation of the known entropies from first principles, and may reveal correction terms. 

Difficulties arising from extending the above simple case to higher dimensions led to the consideration of other kinds of horizon molecules.\cite{Smarr} These difficulties arose as a result of the number of horizon molecules becoming unbounded when the black hole spacetime is unbounded. More complicated definitions of horizon molecules, for example in terms of higher cardinality triads and diamonds, have since also been explored. 

In Ref. \refcite{Barton:2019okw} the original horizon molecule definition was refined and also extended to an entire class of n-element molecules or ``n-molecules''. They succeeded in obtaining well-defined results for higher dimensions, using their definition of molecules. However, they needed to require that $\Sigma$ is strictly spacelike rather than spacelike or null. Ref. \refcite{Machet:2020uml} showed that there are subtleties in the null case that spoil some of the  encouraging results. They also proposed an alternative approach to the problem in terms of the spacetime mutual information (of the action). We refer the reader to Ref. \refcite{Machet:2020uml} for further details on this. 

\section{Closing Remarks\label{sec:conclusion}}
Cosmology offers a rich testing ground for approaches to quantum gravity. The benefits are in both directions, as not only does cosmology benefit from fundamentally motivated answers to open questions, quantum gravity also benefits from the availability of data and observations that potentially bear key signatures of the very largest of energy scales. 

There is no smoking gun evidence yet for any particular quantum gravity effect in cosmological observations. Part of the challenge is that while a quantum theory may suggest an explanation for or even predict an effect, it is not unique in doing so; sometimes, one cannot even rule out a contrived classical explanation for the very same effects. We end with the perspective that if the same theory offers explanations for many different phenomena and open questions in cosmology, it becomes a favorable candidate to underpin them, irrespective of the existence of alternative explanations for  individual questions. 

We have gathered here a brief summary of some work in causal set theory that sheds light on and potentially offers explanations for cosmological open questions such as dark matter and dark energy. It would be valuable to further study and develop each of the directions of research mentioned in this review, and we hope that they will be pursued in the near future.

\section*{Acknowledgments}
 This work has emanated from research conducted with the financial support of Science Foundation Ireland under Grant number 22/PATH-S/10704. I acknowledge financial support from Imperial College London through an Imperial College Research Fellowship grant. This work was also  funded by a Leverhulme Trust Research Project Grant.  
\bibliography{refs}{}
\bibliographystyle{utphys}
\end{document}